\documentclass[aps,pra,reprint,superscriptaddress,showpacs]{revtex4-1}

\usepackage{graphicx}
\usepackage{color}

\newcommand {\bfp} {{\bf p}}
\newcommand {\bfq} {{\bf q}}
\newcommand {\bfr} {{\bf r}}

\renewcommand {\d} {{\rm d}}

\newcommand {\E} {\varepsilon}

\newcommand {\om} {\omega}
\newcommand {\Om} {\Omega}

\begin{document}

\title{Electron production by sensitizing gold nanoparticles irradiated by fast ions}

\author{Alexey V. Verkhovtsev}
\email[]{verkhovtsev@mail.ioffe.ru}
\affiliation{MBN Research Center, Altenh\"oferallee 3, 60438 Frankfurt am Main, Germany}
\affiliation{A.F. Ioffe Physical-Technical Institute, Politekhnicheskaya ul. 26,
194021 St. Petersburg, Russia}

\author{Andrei V. Korol}
\email[]{korol@th.physik.uni-frankfurt.de}
\affiliation{MBN Research Center, Altenh\"oferallee 3, 60438 Frankfurt am Main, Germany}
\affiliation{Department of Physics, St. Petersburg State Maritime Technical University,
Leninskii prospekt 101, 198262 St. Petersburg, Russia}

\author{Andrey V. Solov'yov}
\email[]{solovyov@mbnresearch.com}
\affiliation{MBN Research Center, Altenh\"oferallee 3, 60438 Frankfurt am Main, Germany}
\affiliation{A.F. Ioffe Physical-Technical Institute, Politekhnicheskaya ul. 26,
194021 St. Petersburg, Russia}

%\date{\today}

\begin{abstract}
The yield of electrons generated by gold nanoparticles due to irradiation by fast
charged projectiles is estimated.
%For that, we utilize a model approach based on the plasmon resonance approximation.
%To justify parameters of the model, photoionization spectra of a number of small gold
%nanoparticles are calculated and compared with the spectra obtained by means of
%time-dependent density-functional theory.
The results of calculations are compared to those obtained for pure water medium.
It is demonstrated that a significant increase in the number of emitted electrons
arises from collective electron excitations in the nanoparticle.
The dominating enhancement mechanisms are related to the formation of
(i)~plasmons excited in a whole nanoparticle, and
(ii)~atomic giant resonances due to excitation of $d$~electrons in individual atoms.
%
%It is demonstrated that due to a prominent collective response to an external field,
%noble metal nanoparticles significantly enhance the yield of secondary electrons
%in the medium in the energy range where plasmon excitations, formed in the nanoparticles,
%play a crucial role.
Decay of the collective electron excitations in a nanoparticle embedded in a biological
medium thus represents an important mechanism of the low-energy electron production.
%in the medium.
%Providing this quantitative analysis,
%we indicate the advantage of using gold nanoparticles and in novel techniques of cancer
%therapy with ionizing radiation.
%
Parameters of the utilized model approach are justified through the calculation
of the photoabsorption spectra of several gold nanoparticles, performed by means
of time-dependent density-functional theory.
\end{abstract}

\maketitle

%%%%%%%%%%%%%%%%%%%%%%%%%%%%%%%%%%%%%%%%%%%%%%%%%%%%%%%
\section{Introduction}

In this paper, we perform a theoretical and numerical analysis of electron
production by gold nanoparticles irradiated by fast ions.
It is demonstrated that gold nanoparticles significantly enhance the electron
yield due to the collective response to an external electric field
of a charged projectile.
A significant increase in the number of emitted electrons comes from the two
distinct types of collective electron excitations.
Plasmons, i.e. collective excitations of delocalized valence electrons, dominate
the spectra of electron emission from metallic nanoparticles in the energy range
of about $1-10$~eV.
For higher electron energies (of a few tens of eV), the main contribution
to the electron yield arises from the atomic giant resonance associated with
the collective excitation of $5d$ electrons in individual atoms of a nanoparticle.
As a result of these effects, the number of the low-energy electrons generated
by the gold nanoparticle of a given size significantly exceeds that produced by
an equivalent volume of water.
Thus, decay of the collective electron excitations formed in the nanoparticles
represents an important mechanism of generation of the low-energy electrons.

At present, a vivid scientific interest is in utilizing noble metal
nanoparticles as dose enhancers in cancer treatments with ionizing radiation
\cite{Herold_2000_IntJRadiatBiol.76.1357, Chen_2006_JNanosciNanotechnol.6.1159,
Kobayashi_2010_Mutation_Res.704.123, McMahon_2011_SciRep.1.18}.
Injection of sensitizing nanoparticles into a tumor can increase relative
biological effectiveness of the ionizing radiation, defined as a ratio of the
doses delivered with photons and with a given charged projectile, leading to
the same radiobiological effect.
After the first experimental evidence of radiosensitization by gold nanoparticles
\cite{Hainfeld_2004_PhysMedBiol.49.N309} a number of follow-up experiments with
noble metal and other metallic nanoparticles were performed in recent years
\cite{Porcel_2010_Nanotechnology.21.085103, Kim_2010_Nanotechnology.21.425102,
LeSech_2012_Nanotechnology.23.078001, Liu_2013_Nanoscale.5.11829,
Porcel_2014_NanomedNanotechBiolMed}.

Biodamage due to ionizing radiation involves a number of phenomena, which happen
on various spatial, time and energy scales.
The key phenomena can be described within the so-called multiscale approach to the
physics of radiation damage with ions \cite{Surdutovich_2014_EPJD_Colloquia_Paper}.
In ion-beam cancer therapy %(IBCT)
\cite{Bacarelli_2010_EPJD.60.1, Schardt_2010_RevModPhys.82.383,
Durante_2010_NatRevClinOncol.7.37},
which is one of the promising modern techniques for cancer treatment,
radiation damage is initiated by ions incident on tissue.
Propagating through a biological medium, the projectiles deposit their kinetic
energy by exciting and ionizing the medium.
This interaction leads also to the production of secondary electrons as well as
free radicals and other reactive oxygen species.
It is currently acknowledged that all these secondary species largely cause the
biological damage \cite{Denifl_RadDamage_Chapter, Michael_2000_Science.287.1603,
Surdutovich_2014_EPJD_Colloquia_Paper, Solov'yov_2009_PhysRevE.79.011909}.
%The secondary electrons can directly hit the DNA molecules in the cell nuclei
%or/and generate other secondaries which interact with the DNA.
%
The low-energy electrons, having the kinetic energy from a few eV to several
tens of eV, have been shown to act as important agents of biodamage
\cite{Boudaiffa_2000_Science.287.1658, Huels_2003_JAmChemSoc.125.4467,
Toulemonde_2009_PhysRevE.80.031913}.
In particular, it was indicated that such electrons can produce damage to
biomolecules by dissociative electron attachment \cite{Pan_2003_PhysRevLett.90.208102}.

%In Ref.~\cite{Huels_2003_JAmChemSoc.125.4467, Surdutovich_2012_EPJD.66.206} it has been
%discussed that the Auger effect is one of important channels for production of low-energy
%secondary electrons in biological media, especially in the presence of sensitizing
%nanoparticles \cite{Surdutovich_2014_EPJD_Colloquia_Paper}.
%As a result of recent Monte Carlo simulations \cite{Waelzlein_2014_PhysMedBiol.59.1441},
%a noticeable effect of Auger electrons, emitted from metallic nanoparticles, on the dose
%enhancement has been observed.
%Another important mechanism of production of secondary electrons is associated with the
%interatomic Coulomb decay
%\cite{Cederbaum_1997_PhysRevLett.79.4778, Schwartz_2010_PhysRevLett.105.198102,
%Cederbaum_2013_Nature.505.661}.

It is acknowledged that an important mechanism of excitation/ionization of metallic
clusters and nanoparticles, as well as some other nanoscale systems, relies on the
formation of plasmons,
i.e. collective excitations of delocalized valence electrons that are induced
by an external electric field \cite{Kreibig_Vollmer, Suraud_2013_ClusterScience}.
These excitations appear as prominent resonances in the excitation/ionization spectra
of various systems, and the position of the resonance peak depends strongly on the
type of a system.
In the case of metallic clusters, a typical energy of the plasmon excitations is about
several electronvolts, so the resonance peak is located in the vicinity of the ionization
threshold \cite{Brechignac_1989_ChemPhysLett.164.433, Selby_1989_PhysRevB.40.5417}.
%
%\textcolor{red}{
The effects of resonant enhancement of ionization in atomic (metal, in particular)
clusters exposed to a laser field were reviewed
in Ref.~\cite{Fennel_2010_RevModPhys.82.1793, Wopperer_2014_PhysRep}.
%}

In the recent Monte Carlo simulation \cite{Waelzlein_2014_PhysMedBiol.59.1441},
the authors claimed to include the contribution of plasmon excitations when
calculating the cross sections of electron and proton impact on noble metal
nanoparticles.
However, only ''...the most simplest type of volume plasmon excitation...''
was accounted for in those simulations \cite{Waelzlein_2014_PhysMedBiol.59.1441}.
On this basis, it was stated that the plasmon excitation does not play an
important role in the process of electron emission from metallic nanoparticles,
contributing much less to the overall cross sections than individual excitations.

In this paper, we demonstrate that the decay of plasmon excitations formed in
gold nanoparticles is an important mechanism of generation of the low-energy
secondary electrons.
As we point out, the dominating contribution to the electron yield due to plasmon
excitations comes from the {\it surface} plasmon, since its contribution to the
ionization cross section exceeds that of the volume plasmon, considered in
Ref. \cite{Waelzlein_2014_PhysMedBiol.59.1441}, by an order of magnitude.

Another important mechanism of low-energy electron production by sensitizing
nanoparticles is associated with the collective excitation of $d$ electrons
in individual atoms.
These excitations result in the formation of the so-called atomic giant resonances
in the ionization spectra of many-electron atoms
\cite{Connerade_GR, Brechignac_Connerade_1994_JPhysB.27.3795}.

The results of the present calculations are compared to those carried out for
pure water medium.
We demonstrate that the number of low-energy electrons produced by the gold
nanoparticle of a given size %is more than 40 times higher than
exceeds that produced by an equivalent volume of water by an order of magnitude.
Providing this quantitative analysis, we indicate the advantage of using gold
nanoparticles in cancer treatment with ionizing radiation.
%
%\textcolor{red}{
This result supports the conclusions of the recent experimental studies
\cite{Sanche_2011_Nanotechnology.22.465101, Sanche_2008_RadiatRes.169.19}
which revealed the importance of gold nanoparticles in facilitating the
production of low-energy electrons, which are responsible for DNA damage.
%produced by
%fast charged projectiles interacting with gold nanoparticles, in
%the DNA damage.
%}

The contribution of the plasmon excitations is evaluated by means of
a model approach based on the plasmon resonance approximation (PRA)
\cite{Kreibig_Vollmer, Connerade_AS_PhysRevA.66.013207,
Solovyov_review_2005_IntJModPhys.19.4143, Verkhovtsev_2012_EPJD.66.253}.
Parameters of the utilized model are justified by calculating
photoabsorption spectra of several small gold nanoparticles.
To validate our approach, the PRA-based spectra are compared with those
obtained by means of a more advanced method, namely by time-dependent
density-functional theory (TDDFT) \cite{Runge_Gross_1984_PhysRevLett.52.997}.
To evaluate the contribution of individual atomic excitations, we introduce
an analytical expression, which relates the cross section of photoionization
with that of inelastic scattering in the dipole approximation.
In this paper, we consider gold nanoparticles as a case study but indicate
that the introduced methodology is a general one and can be applied for other
nanoscale systems, which are proposed as sensitizers in ion-beam cancer therapy.

The paper is organized as follows.
Section~\ref{Theory} is devoted to the description of the theoretical framework.
We outline the main points of the TDDFT approach and briefly overview the PRA.
The results of the calculations and their analysis are presented in Section~\ref{Results}.
The photoionization spectra of gold clusters obtained by means of TDDFT are
discussed in Section~\ref{Results_Photo}.
In Section~\ref{Supplementary}, we analyze the nature of the low-energy peak
in the photoabsorption spectra of gold clusters and provide an explanation of
why this feature can be attributed to a collective, plasmon-type excitation of
valence electrons.
In Section~\ref{Results_ElProd}, we analyze the contribution of the plasmon excitations
to the electron production by gold nanoparticles due to irradiation with fast ions.
In Section~\ref{Results_ElProd2}, we account for the individual atomic excitations
and give the resulting quantitative analysis for the electron yield from the
gold nanoparticles.
In Section~\ref{Results_ElProd3}, we consider different kinematic conditions and
estimate the role of the plasmon excitations in the low-energy electron production
as a function of the nanoparticle size and the projectile velocity.
Finally, we draw the conclusions from this work.

The atomic system of units, $m_e = |e| = \hbar = 1$, is used throughout the paper.

%%%%%%%%%%%%%%%%%%%%%%%%%%%%%%%%%%%%%%%%%%%%%%%%%%%%%%%
\section{Theoretical Framework and Computational Details}
\label{Theory}
%%%%%%%%%%%%%%%%%%%%%%%%%%%%%%%%%%%%%%%%%%%%%%%%%%%%%%%

Studying the electron production by gold nanoparticles irradiated by fast
ions, we account for the two collective electron effects, namely excitation
of delocalized valence electrons in a nanoparticle (plasmons) and that of
$d$ electrons in individual atoms (giant resonances).
These phenomena occur in various processes of interaction of ionizing
radiation with matter.
In particular, dipole collective excitations result in the formation of
prominent resonances in the photoabsorption spectra of atomic clusters
and nanoparticles \cite{Kreibig_Vollmer, Suraud_2013_ClusterScience},
while the impact ionization cross sections comprise also the contributions
of higher multipole terms \cite{Solovyov_review_2005_IntJModPhys.19.4143}.
As we demonstrate below, the total photoabsorption spectrum of a gold
nanoparticle in the energy region up to 60~eV is approximately equal
to the sum of the plasmon contribution and that of the $5d$ electron
excitations in individual atoms,
$\sigma_{\gamma} \approx \sigma_{\rm pl} + \sigma_{\rm 5d}$.
Similar to the photoionization, the two distinct types of collective
electron excitations appear in the process of impact ionization.
On this basis, we provide a methodology for analyzing the role of these
contributions to the electron production by sensitizing nanoparticles
separately.

To evaluate the contribution of the plasmon excitations, we have utilized
the PRA approach
\cite{Kreibig_Vollmer, Connerade_AS_PhysRevA.66.013207,
Solovyov_review_2005_IntJModPhys.19.4143, Verkhovtsev_2012_EPJD.66.253}.
It postulates that the dominating contribution to the cross section
comes from collective electron excitations, while single-particle effects
give much smaller contribution in the vicinity of the plasmon resonance
\cite{Gerchikov_1997_JPhysB.30.5939, Gerchikov_2000_PhysRevA.62.043201}.
During the past decades, this approach has provided a clear physical
explanation of the resonant-like structures in photoionization spectra
\cite{Connerade_AS_PhysRevA.66.013207, Verkhovtsev_2013_PhysRevA.88.043201,
Verkhovtsev_2013_JPCS.438.012011}
and differential inelastic scattering cross sections
\cite{Gerchikov_1997_JPhysB.30.4133, Gerchikov_1998_JPhysB.31.3065,
Mikoushkin_1998_PhysRevLett.81.2707, Gerchikov_2000_PhysRevA.62.043201,
Verkhovtsev_2012_JPhysB.45.141002, Bolognesi_2012_EurPhysJD.66.254}
of sodium clusters and carbon fullerenes by the photon and electron impact.
It was also applied
\cite{Connerade_AS_PhysRevA.66.013207, Connerade_AS_1996_JPhysB.29.365,
Gerchikov_1998_JPhysB.31.2331, Korol_AVS_BrS_2014}
to describe the dynamic response of alkali and noble metal clusters in the
processes of radiative electron capture, polarization bremsstrahlung and
multiphoton excitation.

The PRA relies on a few parameters, which include the oscillator strength of
the plasmon excitation, position of the plasmon resonance peak and its width.
The choice of these parameters can be justified by comparing the model-based
cross sections either with experimental data or with the results of more advanced
calculations.
%To the best of our knowledge, there is no experimental data on the impact of
%gold clusters and nanoparticles with photons and charged projectiles for kinetic
%energy of emitted electrons of about $1-10$~eV.
%
%\textcolor{red}{
Ref.~\cite{Sanche_2008_RadiatRes.169.19, Sanche_2011_Nanotechnology.22.465101}
provided an experimental evidence that a considerable portion of radiosensitization
by gold nanoparticles arises from the emitted low-energy electrons.
The experimental yield of $0 - 15$~eV secondary electrons emitted from gold
nanoparticles of about 5~nm in diameter was presented in
Ref.~\cite{Sanche_2011_Nanotechnology.22.465101}.
The cited paper demonstrated an evidence of DNA damage by the low-energy electrons
produced by irradiation of the nanoparticles with fast charged projectiles.
However, to the best of our knowledge, there are no experimentally measured
impact ionization cross sections of gold clusters and nanoparticles with
photons and charged projectiles covering the photon energy / energy loss
range of about $1-10$~eV.
%}
%
Therefore, %in the particular case,
we have calculated photoionization spectra
of several gold clusters by means of TDDFT and then fitted the
{\it ab initio}-based spectra with those calculated within the
model approach.
Such a methodology allowed us to define the resonance frequencies
and to calculate the oscillator strength of the plasmon excitations
in gold nanoparticles.
Note that values of the plasmon width cannot be obtained directly on
the basis of the utilized model.
A precise calculation of the widths can be performed by analyzing the
decay of the collective excitation mode into the incoherent sum of
single-electron excitations.
This process should be considered within the quantum-mechanical
framework \cite{Gerchikov_2000_PhysRevA.62.043201} and cannot be
treated within the classical physics framework, as the PRA does.
Thus, the widths of the plasmon excitations were chosen to obtain
the best agreement with the results of the TDDFT calculations.

\subsection{TDDFT calculations of photoabsorption spectra}
\label{TDDFT_photo}

As a case study we have considered four three-dimensional gold clusters
consisting of 18 to 42 atoms, namely Au$_{18}$, Au$_{20}$, Au$_{32}$,
and Au$_{42}$.
The former two systems, which were observed experimentally, possess
$C_{2v}$ and $T_d$ symmetry, respectively
\cite{Bulusu_2006_PNAS.103.8326, Li_2003_Science.299.864}.
The latter two are hollow, fullerene-like icosahedral structures whose
high stability was predicted theoretically on the basis of DFT calculations
\cite{Johansson_2004_AngewChemIntEd.43.2678, Gu_2004_PhysRevB.70.205401,
Gao_2005_JACS.127.3698}.

\begin{figure}[ht]
\centering
\includegraphics[width=0.41\textwidth,clip]{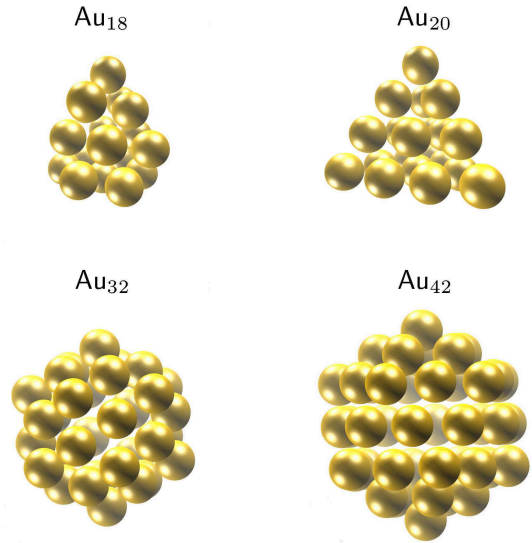}
\caption{Optimized structures of the studied gold clusters.}
\label{fig_Gold_clusters}
\end{figure}

The photoabsorption spectra have been calculated as follows.
At first, we performed the geometry optimization using the Gaussian 09
package \cite{g09}.
The optimization procedure was performed by means of DFT within the
generalized gradient approximation (GGA) and using the effective-core
potential CEP-121G basis set \cite{Stevens_1992_CanJChem.70.612}
augmented by $d$-polarization functions.
The utilized basis set has the $4f$ frozen core, so that 19 electrons
($5s$, $5p$, $5d$, and $6s$) from each gold atom were explicitly treated
in the course of optimization.
To account for the exchange and correlation corrections, the functional
of Perdew, Burke and Ernzerhof \cite{PBE_1996_PhysRevLett.77.3865} was
utilized.
Different spin multiplicities were considered in the course of geometry
optimization.
Figure~\ref{fig_Gold_clusters} shows the ground-state geometries of the
studied clusters.

The linear response function of the clusters was calculated within the dipole approximation.
Within this framework \cite{Kubo_1957_JPhysSocJpn.12.570}, the external potential
acting on a system is represented as a sum of a time-independent part,
$v_{\rm ext}^0(\bfr)$, and a time-dependent perturbation, $v_{\rm ext}^{\prime}(\bfr, t)$.
The time evolution of the electron density, $\rho(\bfr, t)$, is then represented
as a sum of the unperturbed ground-state density and its variation due to
$v^{\prime}_{\rm ext}(\bfr, t)$:
\begin{equation}
\rho(\bfr, t) = \rho_0(\bfr) + \delta\rho(\bfr, t) \ .
\end{equation}

Performing the Fourier transform of the time-dependent quantities involved, one gets
the response of the system to an external perturbation in the frequency representation.
For the perturbation due to a uniform electric field,
$v_{\rm ext}^{\prime}(\bfr,\om) = - {\bf E}(\om) \cdot {\bfr}$,
the Fourier transform of the induced dipole moment reads
\begin{equation}
d_i(\om) = \sum_j \alpha_{ij}(\om) E_j(\om) \ .
\end{equation}

\noindent
Here $i, j$ denote the Cartesian components, and $\alpha_{ij}(\om)$ is the dynamic
polarizability tensor which describes the linear response of the system to the
external electric field:
\begin{equation}
\alpha_{ij}(\om) =
- \int r_i \chi(\bfr,\bfr^{\prime},\om) r_j^{\prime} \, {\rm d}\bfr \, {\rm d}\bfr^{\prime} \ ,
\end{equation}

\noindent
$\chi(\bfr,\bfr^{\prime},\om)$ is the generalized frequency-dependent susceptibility
of the system, and $r_i$ and $r_j^{\prime}$ are the components of the position vectors
$\bfr$ and $\bfr^{\prime}$.
The photoabsorption cross section is related to the imaginary part of the diagonal
elements of the polarizability tensor $\alpha_{ij}(\om)$ through
\begin{equation}
\sigma(\om) = \frac{4\pi\om}{3c} \sum_{j=1}^3 {\rm Im} \, \alpha_{jj}(\om) \ ,
\end{equation}

\noindent where $c$ is the speed of light.

Within the approach introduced in Ref. \cite{Walker_2006_PhysRevLett.96.113001,
Walker_2007_JChemPhys.127.164106, Rocca_2008_JChemPhys.128.154105},
the electron density variation, $\delta\rho(\bfr,\om)$, is expressed via the
so-called Liouvillian operator $\mathcal{L}$,
\begin{equation}
(\om - \mathcal{L}) \cdot \delta\rho(\bfr,\om) =
\left[ v_{\rm ext}^{\prime}(\bfr,\om), \rho_0 \right] \ ,
\end{equation}

\noindent whose action onto $\delta\rho(\bfr,\om)$ is defined as:
\begin{eqnarray}
\mathcal{L} \cdot \delta\rho(\bfr,\om) &=&
\left[ H_0, \delta\rho(\bfr,\om) \right] \nonumber \\ &+&
\left[ v^{\prime}_{\rm H}(\bfr,\om), \rho_0 \right] +
\left[ v^{\prime}_{\rm xc}(\bfr,\om), \rho_0 \right]\ .
\end{eqnarray}

\noindent
Here $H_0$ is the ground-state Kohn-Sham Hamiltonian calculated within
the DFT approach.
The quantities $v^{\prime}_{\rm H}(\bfr,\om)$ and $v^{\prime}_{\rm xc}(\bfr,\om)$
stand for the linear variations of the frequency-dependent electrostatic
and exchange-correlation potentials, respectively \cite{Rocca_2008_JChemPhys.128.154105}.
The polarizability tensor $\alpha_{ij}(\om)$ is defined then by the off-diagonal
matrix element of the resolvent of the Liouvillian $\mathcal{L}$:
\begin{equation}
\alpha_{ij}(\om) =
- \langle r_i | \left( \om - \mathcal{L} \right)^{-1} \cdot [r_j, \rho_0] \rangle \ ,
\end{equation}

\noindent which is calculated using the Lanczos recursion method
(see Ref. \cite{Walker_2006_PhysRevLett.96.113001, Rocca_2008_JChemPhys.128.154105,
Walker_2007_JChemPhys.127.164106} for details).
Based on the frequency representation of the response function, this method
allows one to calculate the total photoabsorption spectrum of a complex
many-electron system in a broad energy range without repeating time-consuming
operations for different excitation energies \cite{Walker_2006_PhysRevLett.96.113001}.

The photoabsorption spectra of the clusters were obtained using the TDDFPT
module \cite{Malcioglu_2011_CompPhysCommun.182.1744} of the Quantum Espresso
software package \cite{Giannozzi_2009_JPhysCondMat.21.395502}.
The optimized geometries were introduced into a supercell of
$20 \times 20 \times 20~{\rm \AA}^3$.
Then, the system of Kohn-Sham equations was solved self-consistently for
valence electrons of the clusters to calculate the ground-state eigenvalues
using a plane-wave approach \cite{Giannozzi_2009_JPhysCondMat.21.395502}.
In the calculations, we used a Vanderbilt ultrasoft nonlinear core-corrected
pseudopotential \cite{Vanderbilt_1990_PhysRevB.41.7892R}, which substitutes
real valence atomic orbitals in the core region with smooth nodeless
pseudo-orbitals \cite{Walker_2007_JChemPhys.127.164106}.
For that, eleven outer-shell electrons ($5d^{10}6s^1$) of each gold atom
were treated as the valence ones.
The obtained results were checked by performing a series of calculations
with different values of the supercell size and the energy cutoff.
The spectra presented in Section~\ref{Results_Photo} were obtained with
the kinetic energy cutoff of 30 Ry for the wave functions and 180 Ry for
the electron densities.

One should note that although the size of the studied systems is not so large
(about 1~nm in diameter),
the largest considered cluster, Au$_{42}$, contains 462 outer-shell electrons,
which should be simultaneously accounted for in the DFT/TDDFT calculations.
This makes the {\it ab initio}-based calculation of the spectra quite demanding
from the computational viewpoint.
To analyze the dynamic response of the systems, which are currently used in the
experimental studies on sensitization
\cite{Porcel_2010_Nanotechnology.21.085103, Kim_2010_Nanotechnology.21.425102,
%McMahon_2011_SciRep.1.18,
LeSech_2012_Nanotechnology.23.078001,
Liu_2013_Nanoscale.5.11829, Porcel_2014_NanomedNanotechBiolMed},
one should therefore utilize model approaches.
They should also describe adequately the response of smaller systems
and must be validated by comparing the results of the {\it ab initio}-
and model-based calculations.

%%%%%%%%%%%%%%%%%%%%%%%%%%%%%%%%%%%%%%%%%%%%%%%%%%%%%%%
\subsection{Plasmon resonance approximation}

To describe plasmon excitations arising in gold nanoparticles, we have adopted
a simple but physically grounded model which treats the studied highly-symmetric
clusters, Au$_{32}$ and Au$_{42}$, as a spherical ''jellium'' shell of a
finite width, $\Delta R = R_2 - R_1$.
In other words, the electron density in these systems is assumed to be homogeneously
distributed over the shell with the thickness $\Delta R$.
%
%Within this model, the studied highly-symmetric clusters, Au$_{32}$ and Au$_{42}$,
%are represented as spherically symmetric objects with a homogeneous charge distribution
%over a full sphere of the radius $R$ (for Au$_{18}$ and Au$_{20}$) or
%over a spherical shell of a finite width, $\Delta R = R_2 - R_1$.
%The charged full sphere representation has been applied earlier for studying electron impact
%ionization of sodium clusters \cite{Gerchikov_1997_JPhysB.30.4133, Gerchikov_1997_JPhysB.30.5939,
%Gerchikov_1998_JPhysB.31.3065, Gerchikov_2000_PhysRevA.62.043201}.
Such a ''jellium''-shell representation has been successfully utilized for
the description of plasmon formation in fullerenes
\cite{Verkhovtsev_2012_JPhysB.45.141002, Oestling_1993_EurophysLett.21.539,
Lambin_Lukas_1992_PhysRevB.46.1794, Lo_2007_JPhysB.40.3973,
Verkhovtsev_2013_PhysRevA.88.043201, Verkhovtsev_2013_JPCS.438.012011}.
In this work, we utilize this model for the description of the fullerene-like
hollow gold clusters.
Values of $\Delta R$ were defined from the analysis of the ground-state
geometries of Au$_{32}$ and Au$_{42}$.
This analysis revealed that the atoms are located on two concentric spheres
of the radii $R_1$ and $R_2$ (see Table~\ref{table_widths}).
%and
%(ii) the radius of metallic bonding for a gold atom is equal to
%1.44~\AA~\cite{Greenwood_ChemElem}.
%These values %which were calculated at the CEP-121G/PBEPBE level of theory
%of $\Delta R$, $R_1$ and $R_2$
%are presented in Table~\ref{table_widths}.
%The parameters $R_1$ and $R_2$ define the radii of the spheres, on which
%gold atoms in the two clusters are located, as calculated at the
%CEP-121G/PBEPBE level of theory.
%The inner and outer radii of the charged shell, $R_1$ and $R_2$, are related to
%the positions of atoms as $R_1 = r_1 - r_{\rm bond}$ and $R_2 = r_2 + r_{\rm bond}$,
%where $r_{\rm bond}$ stands for the radius of metallic bonding.

\begin{table}
\centering
\caption{
Values of $R_1$ and $R_2$ which are used to model the electron
density distribution in hollow Au$_{32}$ and Au$_{42}$ clusters.}
%\begin{tabular}{p{2.5cm}p{1.5cm}p{1.5cm}p{1.5cm}}
\begin{tabular}{p{2.0cm}p{1.5cm}p{1.5cm}}
\hline
%            &  $\Delta R$ (\AA)  &  $R_1$ (\AA) &  $R_2$ (\AA) \\
            &  $R_1$ (\AA) &  $R_2$ (\AA) \\
\hline
%Au$_{32}$   &      0.54    &   3.99  &   4.53  \\
%Au$_{42}$   &      0.65    &   4.60  &   5.25  \\
Au$_{32}$   &      3.99  &   4.53  \\
Au$_{42}$   &      4.60  &   5.25  \\

\hline
\end{tabular}
\label{table_widths}
\end{table}

Within the PRA, the dynamic polarizability $\alpha(\om)$ has a resonance behavior
in the region of frequencies where collective electron modes in a many-electron
system are excited.
Thus, the photoionization cross section $\sigma_{\gamma}(\om)$ of a spherically
symmetric system can be written as:
\begin{equation}
\sigma_{\gamma}(\om) = \frac{4\pi \om}{c} \, {\rm Im} \,\alpha(\om) \sim
{ \om^2 \,\Gamma \over \bigl(\om^2-\om_r^2\bigr)^2+ \om^2 \Gamma^2} \ ,
\label{CS_plasmon}
\end{equation}

\noindent
where $\om$ is the photon energy, $\om_r$ the plasmon resonance frequency,
and $\Gamma$ its width.
The interaction of a hollow system with the uniform external field, $\bf{E}(\om)$,
leads to the variation of the electron density $\delta \rho(\bfr,\om)$ occurring
on the inner and outer surfaces of the hull.
This variation leads to the formation of the surface plasmon, which has two
normal modes, the symmetric and antisymmetric ones
\cite{Oestling_1993_EurophysLett.21.539, Lambin_Lukas_1992_PhysRevB.46.1794,
Lo_2007_JPhysB.40.3973, Korol_AS_2007_PhysRevLett_Comment}.
It has been argued previously
\cite{Connerade_AS_PhysRevA.66.013207, Verkhovtsev_2012_EPJD.66.253,
Korol_AS_2007_PhysRevLett_Comment, Palade_Baran_2014}
that only the surface plasmon can occur in the system interacting with a uniform
external electric field, as it happens in the photoionization process.
When a system interacts with a non-uniform electric field created, for instance,
in collision with a charged particle, the volume plasmon can also occur due to a
local compression of the electron density in the shell interior
\cite{Gerchikov_2000_PhysRevA.62.043201, Verkhovtsev_2012_EPJD.66.253}.
Thus, in the case of irradiation of hollow gold clusters by dipole photons,
two surface plasmon modes, characterized by resonance frequencies $\om_r = \om^{(s)}$
and $\om_r = \om^{(a)}$ and the widths $\Gamma^{(s)}$ and $\Gamma^{(a)}$, are formed.
The frequencies are defined as \cite{Lambin_Lukas_1992_PhysRevB.46.1794,
Oestling_1993_EurophysLett.21.539, Lo_2007_JPhysB.40.3973}:
\begin{equation}
\om^{(s/a)} =
\left[ \frac{N^{(s/a)}}{2(R_2^3 - R_1^3)} \left( 3 \mp p \right) \right]^{1/2} \ ,
\label{eq_01}
\end{equation}

\noindent
where the signs '$-$' and '$+$' correspond to the symmetric $(s)$ and
antisymmetric $(a)$ surface mode, respectively, $p = \sqrt{1 + 8\xi^3}$
with $\xi = R_1/R_2$ being the ratio of the inner to the outer radius.
The values
\begin{equation}
N^{(s)} = N \, \frac{p+1}{2p} \ , \qquad  N^{(a)} = N \, \frac{p-1}{2p}
\end{equation}

\noindent
are the number of delocalized electrons which are involved in each plasmon mode.
They obey the sum rule $N^{(s)} + N^{(a)} = N$ where $N$ stands for a
total number of delocalized electrons in the system.

%Finally, two additional comments should be done.
%First, when a system is treated not a charged spherical layer but as a full sphere of a radius $R$,
%the antisymmetric surface plasmon mode does not contribute to the cross section, so it is defined
%only by a single symmetric mode.
%
%Second,
As indicated above, interaction of the system with a non-uniform electric
field leads also to the formation of the volume plasmon, which appears
due to compression of the electron density inside the volume of the shell
and, therefore, does not interfere with either of the surface plasmon modes.
With neglect of the dispersion, the volume plasmon frequency $\om_p$,
associated with the ground-state electron density $\rho_0$ of $N$ electrons,
is given by
\begin{equation}
\om_p = \sqrt{4 \pi \rho_0} = \sqrt{ 3N /  (R_2^3 - R_1^3) } \ .
\label{eq_02}
\end{equation}

\noindent
The formation of the volume plasmon in the electron impact ionization of
metal clusters and carbon fullerenes was revealed in Ref.
\cite{Gerchikov_2000_PhysRevA.62.043201, Verkhovtsev_2012_JPhysB.45.141002,
%Verkhovtsev_2012_EPJD.66.253,
Bolognesi_2012_EurPhysJD.66.254}.
The model accounting for the contribution of different plasmon modes was
successfully utilized to describe the experimentally observed variation
of the electron energy loss spectra of C$_{60}$ in collision with fast electrons
\cite{Verkhovtsev_2012_JPhysB.45.141002, Bolognesi_2012_EurPhysJD.66.254}.
A detailed explanation of the formation of different plasmon modes can be found
in Ref. \cite{Connerade_AS_PhysRevA.66.013207, Verkhovtsev_2012_EPJD.66.253}.

%%%%%%%%%%%%%%%%%%%%%%%%%%%%%%%%%%%%%%%%%%%%%%%%%%%%%%%
\section{Results and Discussion}
\label{Results}

\subsection{Photoabsorption spectra of gold clusters}
\label{Results_Photo}

\begin{figure}[ht]
\centering
\includegraphics[width=0.45\textwidth,clip]{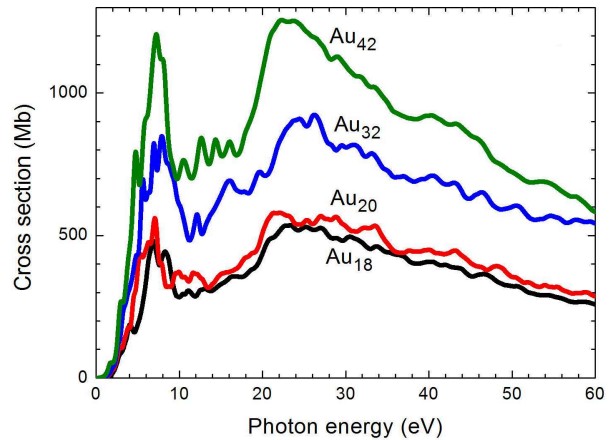}
\caption{
The photoabsorption cross section of the Au$_{18}$ (black), Au$_{20}$ (red),
Au$_{32}$ (blue) and Au$_{42}$ (green) clusters calculated within the TDDFT
framework.}
\label{fig_Gold_clusters_TDDFT}
\end{figure}

\begin{figure}[ht]
\centering
\includegraphics[width=0.46\textwidth,clip]{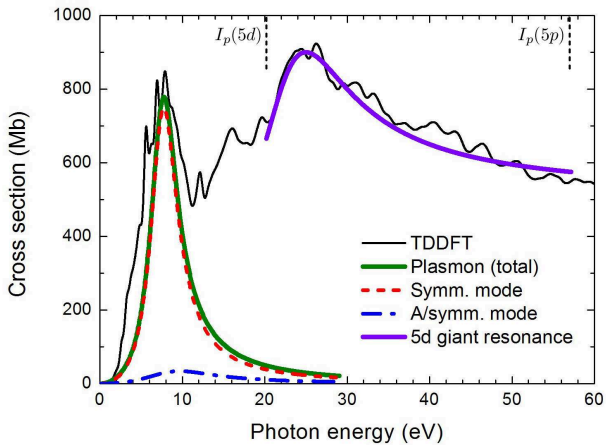}
\caption{
Photoabsorption cross section of Au$_{32}$ calculated within the TDDFT
method (thin black curve).
Thick green curve denotes the contribution of the plasmon excitations
calculated within the PRA.
Red (dashed) and blue (dash-dotted) curves show the contribution of the
symmetric and antisymmetric plasmon modes, respectively.
Thick purple line illustrates the fit of the $5d$ giant atomic resonance
by a Fano-type profile.
Black vertical marks denote the ionization potentials of the $5d$ and $5p$
atomic shells.}
\label{fig_Au32_plasmon}
\end{figure}

The photoabsorption spectra of the Au$_{18}$, Au$_{20}$, Au$_{32}$ and Au$_{42}$
clusters calculated by means of TDDFT for the photon energy up to 60~eV are
presented in Figure~\ref{fig_Gold_clusters_TDDFT}.
The spectra, having a similar profile, are characterized by a low-energy peak
located below 10~eV and by a broad feature which has the maximum at about
$20-25$~eV.
Our analysis has revealed that this feature is the giant resonance formed due to
the excitation of electrons in the $5d$ atomic shell.
To prove this, we have integrated the oscillator strength in the photon energy
range from 20.2~eV (ionization threshold of the $5d$ shell in a single atom of gold)
up to 57.2~eV, which is the ionization threshold of the $5p$ shell
\cite{Henke_1993_AtDataNuclDataTables.54.181}.
The obtained values are 139.8, 153.3, 240.5 and 318.7 for Au$_{18}$, Au$_{20}$,
Au$_{32}$ and Au$_{42}$, respectively.
This indicates that about eight atomic $d$-electrons contribute to the excitation
of the $5d$ shell forming the broad resonance peak in the photoionization spectra.
The $5d$ giant resonance can be fitted with a Fano resonance profile
\cite{Fano_1961_PhysRev.124.1866},
\begin{equation}
\sigma_{5d}(\om) \equiv \sigma_{5d \to \E p, \E f}(\om) \propto
\frac{\left( \Gamma_{5d} + \om - \om_{5d} \right)^2}
{\left( \Gamma_{5d}/2 \right)^2 + \left( \om - \om_{5d} \right)^2} \ ,
\label{Fano}
\end{equation}

\noindent
which is frequently utilized in atomic, nuclear and condensed matter physics
to describe resonant scattering processes occurring in various systems.
Here, $\om_{5d}$ stands for the resonance frequency and $\Gamma_{5d}$ is the
width of the peak.
To describe this feature, we have utilized the values $\om_{5d} = 22$~eV and
$\Gamma_{5d} = 12$~eV.
The fitting for the Au$_{32}$ cluster is illustrated in Figure~\ref{fig_Au32_plasmon}
by the thick solid purple line.
%Note that the Fano profile produces an asymmetric shape of the resonance peak.

The low-energy peak is related to the surface plasmon, which arises due
to collective excitation of delocalized electrons in a whole cluster.
The integration of the oscillator strength in this energy region reveals
that about 1.0 (in Au$_{20}$) to 1.5 (in Au$_{32}$) electrons from each
atom contribute to the plasmon.
Integration of the photoabsorption spectrum of Au$_{32}$ up to 11.2~eV
(energy at which the first dip after the resonance peak is observed)
yields the oscillator strength of 1.43 per atom.
The collective nature of the low-energy peak is analyzed in more detail
in Section~\ref{Supplementary}.

We have fitted the low-energy peak by means of the PRA scheme.
In the case of Au$_{32}$, we have utilized the value $N = 46$, assuming that
1.43 electrons from each of 32 atoms are involved in the plasmon excitation.
The values of $\Gamma^{(s)}$ and $\Gamma^{(a)}$ were chosen to get the best
agreement of the model-based curve with the TDDFT one.
In Figure~\ref{fig_Au32_plasmon}, the thick solid (green) curve represents
the total plasmon contribution to the cross section.
The red and the blue curves illustrate the symmetric and antisymmetric modes,
respectively.
In this calculation, we have utilized the values $\Gamma^{(s)} = 4.0$~eV and
$\Gamma^{(a)} = 10.5$~eV.
The ratio of the widths, $\Gamma^{(s)}/\Gamma^{(a)} = 0.38$, is close to the
value of 0.34, which was utilized for the description of the plasmon excitations
in fullerenes
\cite{Bolognesi_2012_EurPhysJD.66.254, Verkhovtsev_2013_PhysRevA.88.043201}.
Note that the {\it ab initio}-based spectrum reveals a more detailed structure
which is formed atop the plasmon resonance.
This structure represents a series of individual peaks appearing due to
single-particle excitations \cite{Verkhovtsev_2013_PhysRevA.88.043201}
which are neglected in the model.

The oscillator strength, calculated by means of TDDFT in the photon energy range
up to the $5p$ ionization threshold ($\om = 57.2$~eV), is equal to 338.
This value agrees with the total number of valence electrons in the
Au$_{32}$ cluster, $N = 352$, with the relative discrepancy of about 5 percent.
As mentioned above, we assume that about 1.5 and 8 electrons from each atom
contribute to the surface plasmon and the $5d$ giant resonance, respectively.
Thus, we have accounted for the contribution of 9.5 from 11 valence electrons from
each atom.
The contribution of the rest results in a series of individual peaks, positioned in
the photon energy range from 11.2 (the first dip after the low-energy peak) to 20.2~eV
(the $5d$ ionization threshold), that are not accounted for in our model analysis.
Integration of this part of the spectrum yields the oscillator strength of 51.5,
i.e. 1.6 electrons from each atom contribute to the excitations in this energy region.
The individual peaks appear due to single-particle excitations from the $s-d$ band
formed due to hybridization of the $6s$ and $5d$ atomic shells.
On the basis of the performed analysis, the total photoabsorption spectra of gold
clusters in the energy region up to 60~eV can be approximated by the sum of the
plasmon contribution and that of the $5d$ electron excitations in individual atoms,
$\sigma_{\gamma} \approx \sigma_{\rm pl} + \sigma_{\rm 5d}$.

\begin{figure}[ht]
\centering
\includegraphics[width=0.45\textwidth,clip]{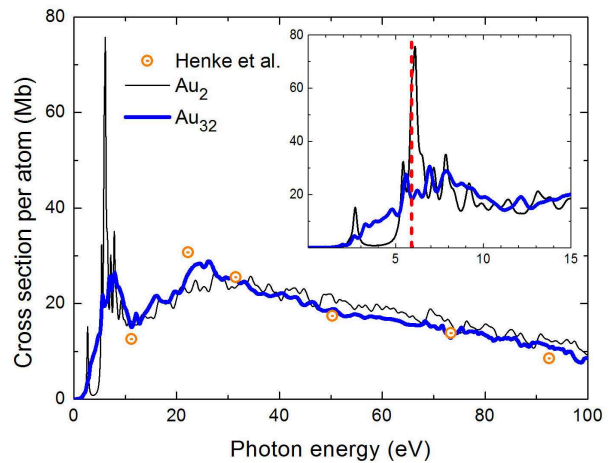}
\caption{
The normalized photoabsorption cross section of Au$_{32}$ (thick blue curve)
and of the Au$_2$ molecule (thin black curve) calculated within the TDDFT framework.
Symbols represent the x-ray absorption data for atomic gold compiled by Henke
{\it et al.} \cite{Henke_1993_AtDataNuclDataTables.54.181}.
The inset shows the low-energy part of the spectra.
Red vertical line illustrates the ionization threshold of a single gold atom
as obtained from the DFT calculations.}
\label{fig_Au2_Au32_Henke}
\end{figure}

In order to stress the different nature of the low- and high-energy features
in the photoabsorption spectra of the clusters, we performed an additional
comparison with the spectrum of a smaller molecular system, a gold dimer.
Figure~\ref{fig_Au2_Au32_Henke} demonstrates the spectra of Au$_{32}$ and Au$_2$
normalized to the number of atoms in each system.
The spectra have a similar profile in the energy region above 15~eV.
This indicates the same origin of excitations in the dimer and in larger atomic clusters,
that is related to the excitation of electrons in the $5d$ shell.
The ripple, which is seen at high photon energies, has an artificial origin
and arises due to the method of calculation of the dipole susceptibility,
which is obtained at complex frequencies with an imaginary part of 0.5~eV.
The normalized spectra are compared also to the x-ray absorption data for atomic
gold taken from the Henke tables \cite{Henke_1993_AtDataNuclDataTables.54.181}.
We state a consistency of the TDDFT-based spectra and well-established set of data.

On the contrary, the low-energy part of the spectrum is quite different
in the two considered cases.
The cross section of the dimer is represented by several well-resolved
narrow peaks due to particular molecular transitions, as opposed to a
broader feature appeared in the spectrum of Au$_{32}$.
The inset of Figure~\ref{fig_Au2_Au32_Henke} shows that the low-energy part
of the spectrum of Au$_2$ is described by the two peaks, positioned at 2.7
and 5.4~eV, followed by the prominent peak at 6.1~eV.
The latter peak corresponds to ionization of the molecule (the ionization
threshold of Au$_2$ is marked by the dashed vertical line).
Therefore, one can observe a clear difference in the structure of the two spectra
below 10~eV, that indicates two different mechanisms of electronic excitations
arising in this energy region, namely well-resolved molecular transitions in the
case of Au$_2$ and the collective excitation of delocalized electrons in the
Au$_{32}$ cluster.
In the following section, we analyze the difference in the structure of the
valence band of Au$_{32}$ and Au$_2$ in greater detail.

\subsection{Plasmon nature of the low-energy peak}
\label{Supplementary}

\begin{figure*}[ht]
\centering
\includegraphics[width=0.98\textwidth,clip]{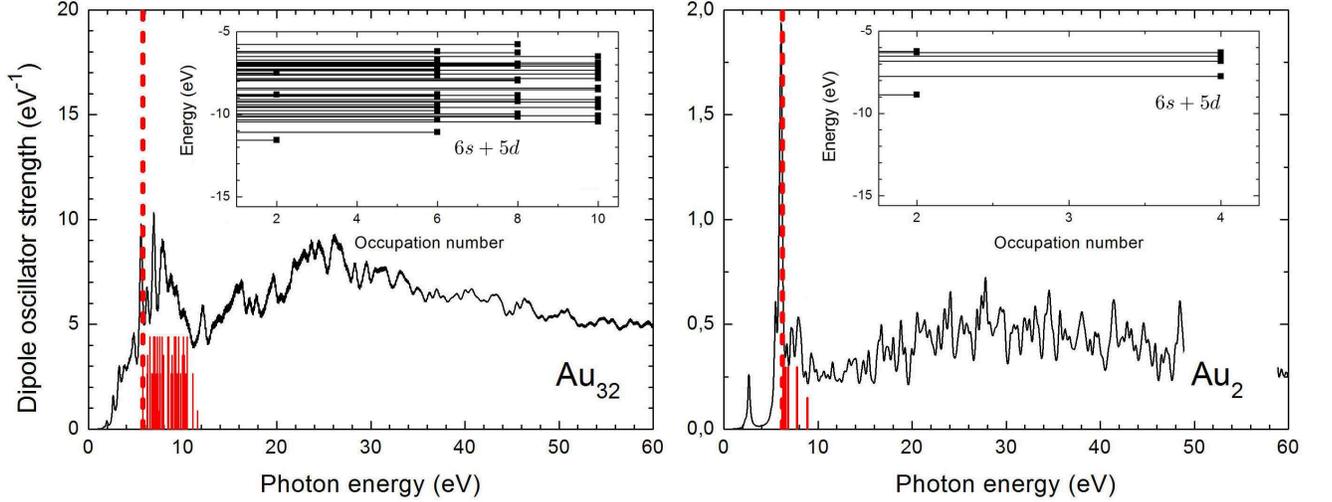}
\caption{
Dipole oscillator strength of the Au$_{32}$ cluster (left) and of
Au$_2$ dimer (right) calculated within the TDDFT framework.
Vertical red lines mark ionization thresholds of each of the $6s + 5d$
valence states.
Thick dashed lines denote the ionization threshold of the HOMO state in
each system.
Inset:
Single-electron energy levels of the valence band as calculated at
CEP-121G(d)/PBEPBE level of theory.
}
\label{Fig_Au32_Au2_valence}
\end{figure*}

Here, we provide an explanation of why we attribute the low-energy
peak in the photoabsorption spectra of gold clusters  to a
%collective,
plasmon-type excitation.
%of delocalized valence electrons.

The term ''plasmon'' is generally used to describe a collective excitation
of delocalized electrons of a system to an external electromagnetic field.
%However, this term is utilized in different scientific communities to describe
%slightly different phenomena which may lead to confusions in its interpretation.
%
Quite commonly, it is understood in terms formulated, for example, by Fano.
In his review \cite{Fano_1992_RevModPhys.64.313} on collective phenomena
in nanoscale systems and in condensed matter, it is stated that
''...common to these phenomena [plasmons, superconductivity, etc., i.e.
those phenomena which are based on the motion of (quasi)independent particles]
is the role of a dense spectrum of states viewed initially as independent.
The seemingly weak interaction among these states often condenses into a
single eigenvalue separated from the rest of the spectrum by an energy gap''.
Thus, if a system has a dense packing of states with a small state separation,
the excitation of these states may be considered as a collective, plasmon-type
one.

In a number of papers
\cite{Malola_2013_ACSNano.7.10263, Philip_2012_NanoLett.12.4661,
Stener_2007_JPhysChemC.111.11862, Durante_2011_JPhysChemC.115.6277},
the term ''surface plasmon resonance'' (SPR) in relation to gold nanoparticles
describes a peak in the visible part of the absoprtion spectra (at about 2.3~eV).
It was stated that this feature is caused by the collective excitation of
$6s$ electrons in the gold atoms
\cite{Malola_2013_ACSNano.7.10263, Philip_2012_NanoLett.12.4661,
Stener_2007_JPhysChemC.111.11862, Durante_2011_JPhysChemC.115.6277}.
The corresponding electronic levels are located in the vicinity of the Fermi
surface, so that these electrons delocalize over the whole nanoparticle.
It was also stated
\cite{Malola_2013_ACSNano.7.10263, Durante_2011_JPhysChemC.115.6277}
that such a SPR (also referred to as the ''localized SPR'', LSPR)
is a characteristic feature of relatively large systems, while the
threshold nanoparticle size for emergence of such plasmonic absorption
is about $1.5-2$~nm (this value corresponds to the number of atoms in
the system of about $150-200$).
Smaller gold nanoparticles (less than 1~nm in diameter) should have
discrete energy levels and, thus, molecular-type transitions between
the occupied and unoccupied states \cite{Philip_2012_NanoLett.12.4661}.

On the other hand, it is also well acknowledged that the occupied $6s$ states
in the gold atoms are strongly hybridized with the $5d$ orbitals.
A general remark on this issue is made in Ref. \cite{Stener_2007_JPhysChemC.111.11862}
that ''...almost always a rather strong mixing is observed, this finding reveals
the ''collective'' nature of the electron excitation''.

In Figure~\ref{Fig_Au32_Au2_valence}, we compare the dipole oscillator
strength distribution of the Au$_{32}$ cluster and of the Au$_2$ molecule.
The low-energy peak in the spectrum of Au$_{32}$ corresponds to the
ionization of the valence band whose structure is shown in the inset.
%
%Figure~\ref{Fig_Au32_Au2_valence} presents the structure of the valence band
%of the Au$_{32}$ cluster compared to the band structure of the Au$_2$ molecule.
The valence orbitals in Au$_{32}$ span over the energy range of about 5.8 eV,
so that the HOMO and the innermost valence state have the ionization potentials
of 5.77 and 11.57~eV, respectively.
The $6s$ and $5d$ orbitals are hybridized and degenerated according to the
cluster symmetry.
Since Au$_{32}$ is a highly-symmetric structure possessing icosahedral symmetry
\cite{Johansson_2004_AngewChemIntEd.43.2678, Gu_2004_PhysRevB.70.205401},
its molecular orbitals are singly, triply, fourfold and fivefold degenerated
in accordance with the irreducible representation of the $I_h$ point group
\cite{Dresselhaus_GroupTheory}.
In Au$_{32}$, there are 176 valence orbitals which are reduced to 46 because
of symmetry of the cluster.
Analysis of the valence state separation indicates that this value varies from
0.01 to 0.64~eV, resulting in the average value of 0.13~eV.
This value corresponds to those calculated \cite{Aikens_2008_JPhysChemC.112.11272}
for a number of silver clusters Ag$_n$ ($n = 20 \dots 120$).
The emergence of a plasmon peak was observed in these systems by means of the
TDDFT approach.
%With increasing the cluster size.

Based on the analysis of the valence band structure and on the explanation
given by Fano \cite{Fano_1992_RevModPhys.64.313}, we thus associate
the low-energy peak in the photoabsorption spectra of the gold clusters with
a collective, plasmon-type excitation.
However, we note that this excitation involves not only the $6s$ electrons
(as it happens in the case of the LSPR located at about 2.3~eV
\cite{Malola_2013_ACSNano.7.10263, Philip_2012_NanoLett.12.4661,
Stener_2007_JPhysChemC.111.11862, Durante_2011_JPhysChemC.115.6277})
but also some fraction of the $5d$ electrons because of the strong
overlap between the $s$ and $d$ states.

To the further support of this statement, we have analyzed this collective
excitation in terms of the classical Mie theory.
The Au$_{32}$ cluster has a hollow, fullerene-like structure
%\cite{Johansson_2004_AngewChemIntEd.43.2678, Gu_2004_PhysRevB.70.205401},
so that the electron density is assumed to be homogeneously distributed
in between the two spheres on which the atoms are positioned.
Calculating the frequency of a surface plasmon excited in a hull-like
system with the help of Eq.~(\ref{eq_01}), one derives the value of 6.3~eV
that matches well the position of the dominant feature in the TDDFT-based
spectrum of Au$_{32}$.
Thus, we can state that the low-energy peak arises due to the collective,
plasmon-type excitation of electrons delocalized over the whole cluster.
In this estimate, we assumed that 46 electrons (1.43 electrons from each atom)
delocalize and participate in the collective excitation.
This value is in agreement with the results of
Ref. \cite{Fernandez_2006_PhysRevB.73.235433}, where the dipole polarizability
of a series of three-dimensional gold clusters as a function of their size was
calculated.
It was found that the calculated polarizabilities suggest a delocalized character
of some fraction of $d$ electrons, so that 1.56 delocalized valence electrons
contribute to the linear response to an external field.

\subsection{Electron production via the plasmon excitation mechanism}
\label{Results_ElProd}

Having justified the parameters of the model that describe the dipole plasmon
excitation in gold nanoparticles, we extend the model approach to study the
electron production due to the plasmon excitation mechanism.
Within the PRA, the double differential inelastic scattering cross section of
a fast projectile in collision with a hull-like system can be defined as a sum
of three terms \cite{Verkhovtsev_2012_EPJD.66.253, Verkhovtsev_2012_JPhysB.45.141002}:
\begin{equation}
\frac{\d^2\sigma_{\rm pl}}{\d\E_2 \d\Om_{{\bfp}_2}} =
\frac{\d^2\sigma^{(s)} }{\d\E_2 \d\Om_{{\bfp}_2}} +
\frac{\d^2\sigma^{(a)} }{\d\E_2 \d\Om_{{\bfp}_2}} +
\frac{\d^2\sigma^{(v)} }{\d\E_2 \d\Om_{{\bfp}_2}} \ ,
\label{Equation.01}
\end{equation}

\noindent
which describe the partial contribution of the surface
(the two modes, $s$ and $a$) and the volume ($v$) plasmons.
Here $\E_2$ is the kinetic energy of the scattered projectile,
${\bfp}_2$ its momentum, and $\Om_{{\bfp}_2}$ its solid angle.
The cross section $\d^2\sigma_{\rm pl} / \d\E_2 \d\Om_{{\bfp}_2}$ can be written
in terms of the energy loss $\Delta \varepsilon = \E_1 - \E_2$ of the incident
projectile of energy $\E_1$.
Integration of $\d^2\sigma_{\rm pl} / \d\Delta\E \, \d\Om_{{\bfp}_2}$ over the
solid angle leads to the single differential cross section:
\begin{equation}
\frac{{\rm d}\sigma_{\rm pl}}{{\rm d}\Delta\E} =
\int {\rm d}\Omega_{{\bf p}_2}
\frac{{\rm d}^2\sigma_{\rm pl}}{{\rm d}\Delta \varepsilon \, {\rm d}\Omega_{{\bf p}_2}}
= \frac{2\pi}{p_1 p_2} \int\limits_{q_{\rm min}}^{q_{\rm max}} q \, {\rm d}q
\frac{{\rm d}^2\sigma_{\rm pl}}{{\rm d}\Delta \varepsilon \, {\rm d}\Omega_{{\bf p}_2}} \ ,
\end{equation}

\noindent
where
%$\Delta \varepsilon = \E_1 - \E_2$ is the energy loss,
%$\E_1 = {\bfp}_1^2/2$ the kinetic energy of the incident projectile with
${\bfp}_1$ is the initial momentum of the projectile and
${\bfq} = {\bfp}_1 - {\bfp}_2$ the transferred momentum.
Explicit expressions for the contributions of the surface and the volume
plasmons, entering Eq.~(\ref{Equation.01}), obtained within the plane-wave
Born approximation, are presented in Ref. \cite{Verkhovtsev_2012_EPJD.66.253}.
The Born approximation is applicable since the considered collision velocities
($v = 2 - 20$~a.u.) are significantly larger than the characteristic velocities
of delocalized electrons in the gold nanoparticles ($v \approx 0.5$~a.u.).

The surface and the volume plasmon terms appearing on the right-hand side of
Eq.~(\ref{Equation.01}) are constructed as a sum over different multipole
contributions corresponding to different values of the angular momentum $l$:
\begin{eqnarray}
\begin{array}{l l}
\displaystyle{ \frac{\d^2\sigma^{(i)} }{\d\E_2 \d\Om_{{\bfp}_2}} }
\propto
%\sum\limits_{l=0}^{\infty} \frac{ \om_{l}^{(i)2}\, \Gamma_{l}^{(i)}  }
\sum\limits_{l} \frac{ \om_{l}^{(i)2}\, \Gamma_{l}^{(i)}  }
{ \bigl(\om^2-\om_{l}^{(i)2}\bigr)^2 + \om^2\Gamma_{l}^{(i)2} }
\vspace{0.2cm} \\
\displaystyle{ \frac{\d^2\sigma^{(v)} }{\d\E_2 \d\Om_{{\bfp}_2}} }
\propto
%\sum\limits_{l=0}^{\infty} \frac{ \om_p^2\, \Gamma_l^{(v)} }
\sum\limits_{l} \frac{ \om_p^2\, \Gamma_l^{(v)} }
{ \bigl(\om^2-\om_p^2\bigr)^2+\om^2\Gamma_l^{(v)2} } \ ,
\end{array}
\label{Equation.02}
\end{eqnarray}

\noindent
where $i = s,a$ denotes the two modes of the surface plasmon.
The frequencies of the symmetric and antisymmetric modes of the multipolarity $l$
are given by \cite{Verkhovtsev_2012_EPJD.66.253, Oestling_1993_EurophysLett.21.539}:
\begin{equation}
\displaystyle{
\om_{l}^{(s/a)} =
\left( 1 \mp  \frac{1}{2l+1} \sqrt{1 + 4l(l+1)\xi^{2l+1}} \right)^{1/2}
\frac{\om_p}{\sqrt{2}}
}
\label{MultipoleVariation.2a}
\end{equation}

\noindent
where '$-$' and '$+$' stand for symmetric $(s)$ and antisymmetric $(a)$ modes,
respectively.
In the dipole case ($l=1$), this expression reduces to Eq.~(\ref{eq_01}).
The volume plasmon frequency, $\om_p$, is independent on $l$ as it follows
from Eq.~(\ref{eq_02}).

\begin{figure}[ht]
\centering
\includegraphics[width=0.47\textwidth,clip]{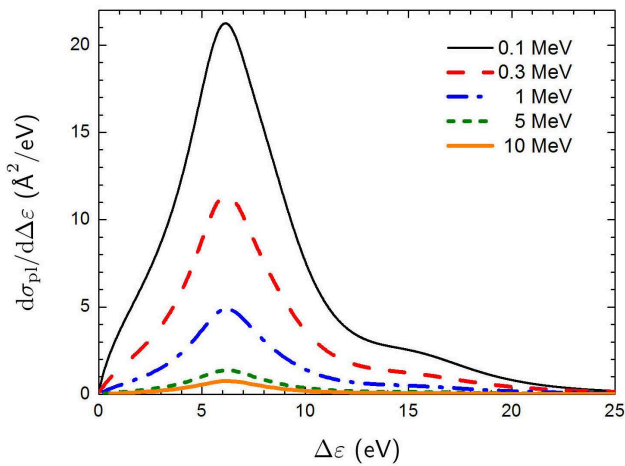}
\caption{Contribution of the plasmon excitations to the single differential
cross section, ${\rm d}\sigma_{\rm pl}/{\rm d}\Delta\E$, of the Au$_{32}$
cluster irradiated by fast protons of different incident energies as a
function of the energy loss.}
\label{fig_Au32_SDCS}
\end{figure}

Figure~\ref{fig_Au32_SDCS} shows the cross section
${\rm d}\sigma_{\rm pl}/{\rm d}\Delta\E$ calculated for the Au$_{32}$ cluster
irradiated by fast protons of different incident energies as indicated.
%Note that the incident energy of about 0.3~MeV per nucleon corresponds
%to the energy of an ion in the vicinity of the Bragg peak
%\cite{Surdutovich_2007_EPJD.51.63, Surdutovich_2010_PhysRevE.82.051915}.
The figure demonstrates that the amplitude and the shape of the plasmon
resonance depend strongly on the kinetic energy of the proton.
It was shown previously \cite{Gerchikov_1997_JPhysB.30.4133} that the
relative contribution of the quadrupole $(l=2)$ and higher multipole
terms to the cross section decreases significantly with an increase of
the collision velocity.
At high velocities, the dipole contribution dominates over the higher
multipole contributions, since the dipole potential decreases slower
at large distances than the higher multipole potentials.
%
%higher multipole contributions become significant only in the case when the characteristic
%collision distances, $v / \om_r$, become comparable with the cluster size.
%In the opposite case of large collision distances, $v / \om_r \ll R$, the dipole contribution
%dominates over the higher multipole contributions, since the dipole potential decreases
%more slowly at large distances than the higher multipole potentials.
%The shape of the differential energy loss spectrum depends on the collision velocity, because
%the number of multipoles contributing significantly to the cross section is different in various
%collision energy ranges.
%This figure shows that the dipole resonance becomes dominant as the collision velocity increases.

The presented spectra comprise contributions of both the surface and the
volume plasmon excitations, and different multipole terms contribute to
each of them.
Calculating the cross sections presented in Figure~\ref{fig_Au32_SDCS},
we accounted for the contribution of the dipole ($l= 1$), quadrupole
($l = 2$) and octupole ($l = 3$) terms.
The excitations with large angular momenta have a single-particle rather than
a collective nature \cite{Gerchikov_1997_JPhysB.30.4133}.
With increasing $l$, the wavelength of a plasmon excitation, $\lambda_{\rm pl} = 2\pi R/l$,
becomes smaller than the characteristic wavelength of the delocalized electrons
in the system, $\lambda_e = 2\pi /\sqrt{2\epsilon}$.
Here $\epsilon \sim I_p$ is the characteristic electron excitation energy in
the cluster, and $I_p$ is the ionization threshold of the system.
The DFT calculations of the electronic structure of Au$_{32}$ derived the value
of $I_p = 5.77$~eV, while the calculated HOMO-LUMO gap is 1.54~eV.
Thus, one estimates a characteristic excitation energy of delocalized electrons
in Au$_{32}$ to be of the order of several electronvolts, that results in the
account of the three multipole plasmon terms.

Following the methodology utilized in Ref. \cite{Bolognesi_2012_EurPhysJD.66.254},
we assume that the ratio $\gamma_l = \Gamma_l/\om_l$ of the width of the plasmon
resonance to its frequency equals to $\gamma_l^{(s)} = 0.6$ for all multipole terms
of the symmetric mode, and to $\gamma_l^{(a)} = 1.0$ for the antisymmetric mode.
In Ref. \cite{Verkhovtsev_2013_PhysRevA.88.043201, Bolognesi_2012_EurPhysJD.66.254}
these values were successfully utilized to describe the main features
of the photon and electron impact ionization cross sections of the C$_{60}$ fullerene
whose topology is similar to the Au$_{32}$ cluster.
For the volume plasmon we consider the ratio $\gamma_l^{(v)} = \Gamma_l^{(v)}/\om_p = 1.0$.

\begin{figure}[ht]
\centering
\includegraphics[width=0.47\textwidth,clip]{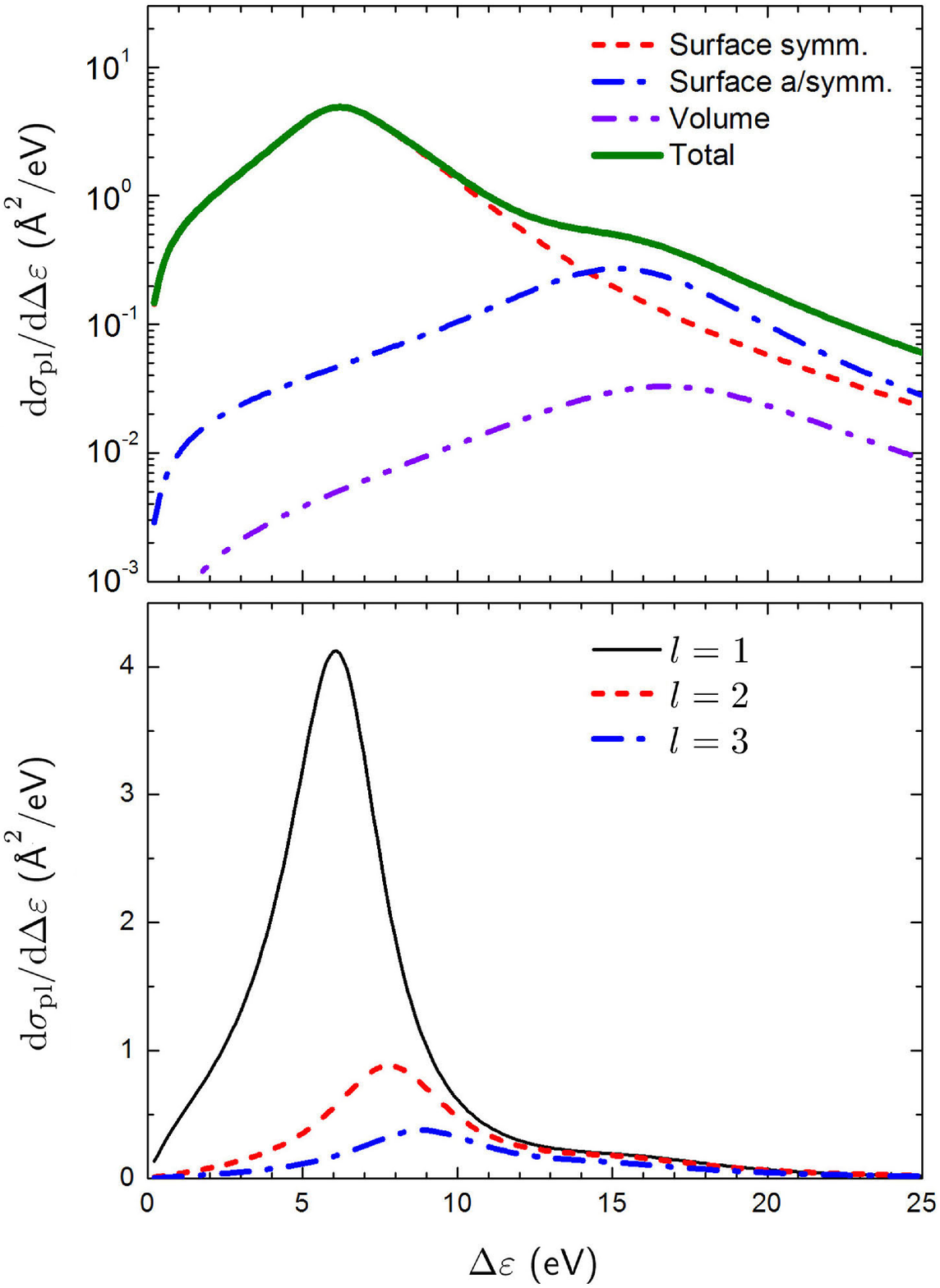}
\caption{
Single differential cross section ${\rm d}\sigma_{\rm pl}/{\rm d}\Delta\E$
of the Au$_{32}$ cluster irradiated by a 1 MeV proton as a function of the
energy loss.
Upper panel illustrates the contribution of different plasmon excitations to
the resulting spectrum.
Lower panel shows the contribution of different multipole terms.}
\label{fig_Au32_SDCS_2}
\end{figure}

Figure~\ref{fig_Au32_SDCS_2} illustrates the contribution of different plasmon
modes to the spectrum of Au$_{32}$ irradiated by a 1 MeV proton and also partial
contributions of different multipole modes.
The main contribution to the cross section comes from the symmetric mode of the
surface plasmon, which, in turn, is dominated by the dipole excitation.
The figure shows that the relative contribution of the surface plasmon exceeds
that of the volume plasmon by more than an order of magnitude.
Thus, the leading mechanism of electron production by gold nanoparticles is
related to the surface plasmon.
This result contradicts with the recent Monte Carlo simulations
\cite{Waelzlein_2014_PhysMedBiol.59.1441}, which claimed that the plasmon excitations
do not play an important role in the process of electron emission from metallic
nanoparticles.
Let us stress that only the volume plasmon excitation was accounted for in those simulations.
Below we demonstrate that the emission of the low-energy electrons from the gold nanoparticles
is indeed a prominent effect, which should be accounted for when estimating the secondary
electron yield in a biological medium with embedded nanoparticles.

%%%%%%%%%%%%%%%
Note that the maximum of the resonance peak is located at 6.3~eV,
that is slightly above the ionization threshold of Au$_{32}$,
$I_p = 5.77$~eV.
The plasmons located above the ionization threshold can decay via
the ionization process \cite{Gerchikov_2000_PhysRevA.62.043201}.
On the contrary, the decay of a collective excitation located below the
ionization threshold results in single-electron excitations, which can
also be coupled with the ionic motion by the electron-phonon coupling
\cite{Gerchikov_2000_JPhysB.33.4905}.
%
%In general, the width of plasmon resonance appears via the decay of the collective
%excitations to the single-particle ones.
%In the case of damping of the plasmons located below the ionization threshold,
%single-electron excitations in the region of the plasmon resonance have the
%discrete spectrum.
%However, discrete excitations can be broadened due to the electron-ion coupling
%\cite{Gerchikov_2000_JPhysB.33.4905}.
Therefore, decay of the surface plasmon in Au$_{32}$ results in the electron
emission from the system which can be accompanied by vibrations of the ionic core.
%%%%%%%%%%%%%%%

\begin{figure}[ht]
\centering
\includegraphics[width=0.47\textwidth,clip]{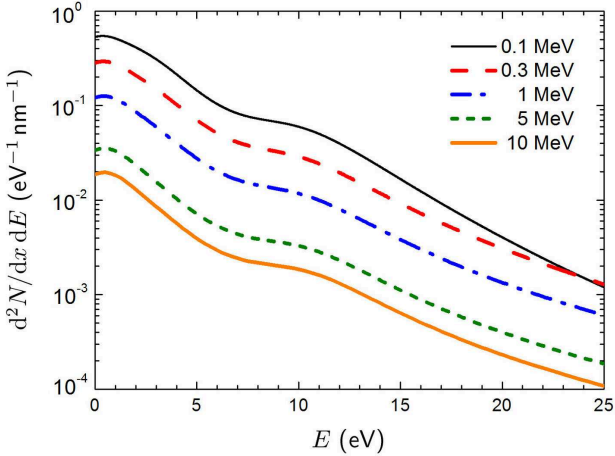}
\caption{Number of electrons per unit length per unit energy emitted via the
plasmon excitation mechanism from the Au$_{32}$ cluster irradiated by a fast proton.
Different curves illustrate different values of the proton's kinetic energy.}
\label{fig_Au_ElProd}
\end{figure}

\begin{figure}[ht]
\centering
\includegraphics[width=0.47\textwidth,clip]{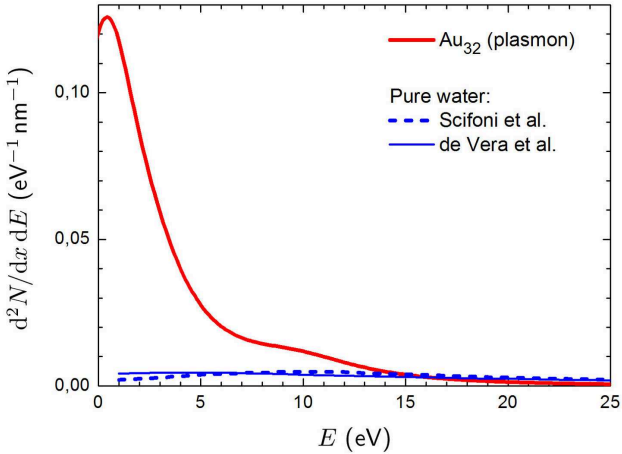}
\caption{
Number of electrons per unit length per unit energy emitted via the plasmon
excitation mechanism from the Au$_{32}$ cluster irradiated by a 1~MeV proton
(red curve).
Blue curves represent the number of electron generated from an equivalent
volume of pure water.
Solid and dashed blue curves represent the results obtained within the dielectric
formalism by Scifoni {\it et al.} \cite{Scifoni_2010_PhysRevE.81.021903} and
de Vera {\it et al.} \cite{deVera_2013_PhysRevLett.110.148104}, respectively.}
\label{fig_Au_ElProd_2a}
\end{figure}

To quantify the production of electrons via the plasmon excitation mechanism,
we redefine the cross section ${\rm d}\sigma_{\rm pl} / {\rm d}\Delta\E$ as a
function of the kinetic energy $E$ of emitted electrons.
This quantity is related to the energy loss via $E = \Delta\E - I_p$, where
$I_p$ is the ionization threshold of the system.
The cross section ${\rm d}\sigma_{\rm pl} / {\rm d}E$ can be related to the
probability to produce $N$ electrons with kinetic energy $E$, in the interval $dE$,
emitted from a segment ${\rm d}x$, via %of the trajectory of a single ion
\cite{Surdutovich_2014_EPJD_Colloquia_Paper}:
\begin{equation}
\frac{{\rm d}^2 N(E)}{{\rm d}x \,{\rm d}E} =
\frac{1}{V} \frac{{\rm d}\sigma_{\rm pl}}{{\rm d}E} \ ,
\end{equation}

\noindent
where $V$ is the volume occupied by the nanoparticle.
Assuming that the linear size of the Au$_{32}$ cluster is $d \approx 0.9$~nm,
we have calculated the number of electrons per unit length per unit energy
emitted from this system due to proton irradiation.
Figure \ref{fig_Au_ElProd} illustrates the dependence of this quantity on
kinetic energy of emitted electrons for different incident energies of the proton.

In Figure~\ref{fig_Au_ElProd_2a}, we compare the electron production
by Au$_{32}$ (red curve) and by an equivalent volume
of pure water medium (blue curves) irradiated by a 1~MeV proton.
Solid and dashed blue curves represent the results obtained recently
within the dielectric formalism
\cite{Scifoni_2010_PhysRevE.81.021903, deVera_2013_PhysRevLett.110.148104}.
This approach is based on the experimental measurements of the energy-loss
function of the target medium, ${\rm Im}[-1/\epsilon(\om, q)]$, where
$\epsilon(\om, q)$ is the complex dielectric function, with $\om$ and
$q$ being the energy and the momentum transferred to the electronic
excitation, respectively.
Comparative analysis of the spectra reveals that the number of the
low-energy electrons (with the kinetic energy of about a few eV)
produced by the gold nanoparticle is about one order of magnitude
higher than by liquid water.

In the above-presented analysis, we have considered the hollow cluster
of diameter $d \approx 0.9$~nm.
An additional estimate was done also for a solid nanoparticle of a similar size.
As a case study, we considered a 1~nm nanoparticle ''cut'' from an ideal
gold crystal having the face-centered cubic (fcc) lattice with the parameter
$a = 4.08$~\AA.
The crystalline structure was constructed using the ''Crystal generator''
tool \cite{Sushko_2013_JComputPhys.252.404} of the MBN Explorer
\cite{MBN_Explorer} software.
As a result, we found that the small solid gold nanoparticle is composed of 31 atoms.
Thus, its atomic density is close to that of the above-considered Au$_{32}$ cluster.

The dynamic response of a solid nanoparticle can also be modeled by means
of the PRA formalism assuming that the system is treated not as a ''jellium''
hull but as a full sphere, where the electron density is uniformly distributed
inside the sphere of a radius $R$
\cite{Gerchikov_1997_JPhysB.30.5939, Gerchikov_2000_PhysRevA.62.043201}.
In this case, the electron density variation on the surface and in the volume of
the nanoparticle leads to the formation of the surface (symmetric mode) and the
volume plasmon, respectively, while the antisymmetric surface plasmon mode does
not contribute to the cross section (a detailed explanation of this phenomenon
can be found in Ref.~\cite{Verkhovtsev_2012_EPJD.66.253}).
As shown in Figure~\ref{fig_Au32_SDCS_2} in the case of the hollow system,
the contribution of the antisymmetric mode in the $1 - 10$~eV range
is an order of magnitude smaller than that of the symmetric mode.
Thus, the absence of the antisymmetric mode in the excitation spectrum of the
solid nanoparticle should not lead to quantitatively different results from
those presented in Figure~\ref{fig_Au32_SDCS_2}.
Therefore, all the above-given estimates on the electron production yield by
the hollow nanoparticle of about 1~nm diameter should also be valid for the
case of the solid fcc structure.

\subsection{Contribution of individual atomic excitations to electron production}
\label{Results_ElProd2}

As illustrated above (see Figure~\ref{fig_Au32_plasmon}), there is also a prominent
contribution of the atomic $5d$ electrons to the ionization cross section.
The $d$ electrons in the atoms of gold play a dominant role at the excitation energies
from approximately 20 to 60~eV.
%
%For distant collisions, the ionization spectra are dominated by the dipole term.
For distant collisions, i.e. when the impact parameter exceeds the radius $R_{\rm at}$
of the atomic subshell, the ionization spectra are dominated by the dipole term
\cite{Landau_Lifshitz_3}.
On this basis, we have compared the cross sections of photoionization,
$\sigma_{\gamma}$, and the dipole term of atomic inelastic scattering,
${\rm d}\sigma_{\rm 5d} / {\rm d}\Delta\E$, calculated in the Born approximation.
%On this basis, we have estimated the atomic inelastic scattering cross section
%${\rm d}\sigma_{\rm 5d} / {\rm d}\Delta\E$ in the dipole approximation,
As a result, one derives the following expression:
\begin{equation}
\frac{{\rm d}\sigma_{5d}}{{\rm d}\Delta\E} =
\frac{2c}{\pi \om v_1^2} \sigma_{\gamma} \ln{ \left( \frac{v_1}{\om R_{5d}} \right)} \ ,
\label{ElScatter_to_PI}
\end{equation}

\noindent
where $\sigma_{\gamma} \equiv \sigma_{5d}(\om)$ is the $5d$ photoionization
cross section estimated by a Fano-type profile, Eq.~(\ref{Fano}),
$\om = \E_1 - \E_2$ the energy transfer,
$v_1$ the projectile velocity,
and $R_{5d}$ a characteristic radius of the $5d$ electron shell.
Equation~(\ref{ElScatter_to_PI}), obtained within the so-called
''logarithmic approximation'', assumes that the main contribution to
the cross section ${\rm d}\sigma_{\rm 5d} / {\rm d}\Delta\E$ comes from
the region of large distances, $R_{5d} < r < v_1/\om$.
This relation has the logarithmic accuracy which implies that the
logarithmic term dominates the cross section while all non-logarithmic
terms are neglected \cite{Korol_AVS_BrS_2014}.
Making an estimate for the gold atoms, we assumed $\om \approx 1$~a.u.
which corresponds to the maximum of the $5d$ giant resonance in gold
\cite{Henke_1993_AtDataNuclDataTables.54.181},
$v_1 \approx 6.3$~a.u. for a 1~MeV proton,
and the electron shell radius $R_{5d}({\rm Au}) \approx$~2~a.u.
Note that the interaction of the incident projectile with the nanoparticle
leads to the formation of the $5d$ giant resonance not in all atoms of the
system but only in those located within the impact parameter interval from
$r_{\rm min} \simeq R_{5d}$ to $r_{\rm max} \simeq v_1/\om$.
This estimate indicates that the $5d$ giant resonance is formed in about
one third of atoms of the nanoparticle.

The number of electrons per unit length per unit energy produced via the
excitation of $5d$ electrons in individual gold atoms, is defined as:
\begin{equation}
\frac{{\rm d}^2 N(E)}{{\rm d}x \,{\rm d}E} =
A \, n \,  \frac{{\rm d}\sigma_{5d}}{{\rm d}E} \ ,
\label{ElProd_5d}
\end{equation}

\noindent
where $n$ is the atomic density of the target, and
$A$ the ratio of the number of atoms possessing the $5d$ resonance to the
total number of atoms in the nanoparticle.

\begin{figure}[ht]
\centering
\includegraphics[width=0.45\textwidth,clip]{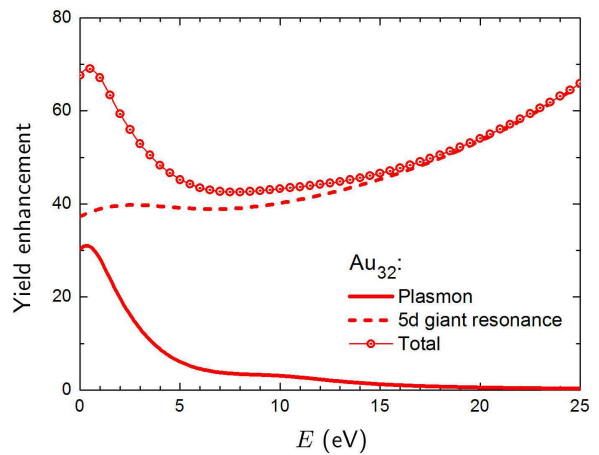}
\caption{
Yield enhancement from the Au$_{32}$ cluster as compared to an equivalent
volume of pure water \cite{deVera_2013_PhysRevLett.110.148104}.
The solid and dashed lines show the contribution of the plasmons and the
atomic $5d$ excitations, respectively.
Symbols illustrate the resulting enhancement.}
\label{fig_Au_ElProd_2b}
\end{figure}

To estimate the total number of electrons produced due to the
collective excitations in the gold nanoparticle, we have
accounted for the contribution of the plasmon excitations
and excitations of $5d$ electrons in individual atoms.
Figure~\ref{fig_Au_ElProd_2b} shows the relative enhancement
of the electron yield from the Au$_{32}$ cluster as compared
to pure water.
The data for the gold nanoparticle are normalized to the spectrum
for liquid water \cite{deVera_2013_PhysRevLett.110.148104}.
The solid line shows the contribution of the plasmon excitations
to the electron yield, while the dashed line presents the contribution
from the atomic $5d$ giant resonance, estimated using
Eq.~(\ref{ElScatter_to_PI}) and (\ref{ElProd_5d}).
Making this estimate, we have assumed that the ionization cross
sections of individual atoms are dominated by the dipole excitation.
Contribution of quadrupole and higher multipole terms will lead to
an increase in the number of emitted electrons but their relative
contribution will be not as large as that from the dipole excitation.
Accounting for the plasmon contribution leads to a significant additional
increase in the number of $1-5$~eV emitted electrons as compared to the
pure water.
Due to the collective electron excitations arising in the $\sim 1$~nm gold
nanoparticle, it can thus produce up to 70 times larger number of the
low-energy electrons comparing to the equivalent volume of pure water medium.
The enhancement of the secondary electron yield may increase the probability
of the tumor cell destruction due to the double- or multiple strand break of
the DNA \cite{Surdutovich_2014_EPJD_Colloquia_Paper}.
The results of the performed analysis indicate that the decay of the collective
electron excitations in gold nanoparticles is an important mechanism of the
yield enhancement.

\subsection{Different kinematic conditions}
\label{Results_ElProd3}

In this section, we analyze how the contribution of the plasmon and the
$5d$ excitation mechanisms evolve at different kinematic conditions,
namely for different projectile velocities and for the nanoparticles of
different size.

\begin{figure}[ht]
\centering
\includegraphics[width=0.47\textwidth,clip]{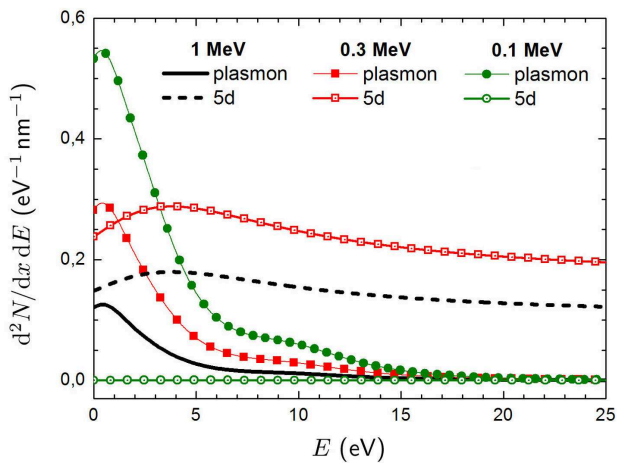}
\caption{
Number of electrons per unit length per unit energy produced via the plasmon
and the $5d$ excitation mechanisms in the Au$_{32}$ cluster irradiated by a
proton of different kinetic energies.
Solid line and filled symbols illustrate the plasmon contribution to the electron
production yield.
Dashed line and open symbols show the contribution of the $5d$ giant resonance.}
\label{fig_Au_ElProd_3}
\end{figure}

In the case of the Au$_{32}$ cluster irradiated by a 1~MeV proton, the
number of electrons produced via the excitations in individual atoms is
generally higher than that produced via the plasmon excitation mechanism
(see Figure \ref{fig_Au_ElProd_2b}).
At certain kinematic conditions, the plasmon contribution to the low-energy
(of about $1-10$~eV) electron yield from the gold nanoparticle can exceed
significantly that due to the atomic giant resonance.
An illustration of this effect is given in Figure~\ref{fig_Au_ElProd_3},
where we compare the yield of electron production due to irradiation by 1
and 0.1~MeV protons.
We also consider the incident energy of 0.3~MeV which is approximately equal
to that of an ion in the vicinity of the Bragg peak
\cite{Surdutovich_2007_EPJD.51.63, Surdutovich_2010_PhysRevE.82.051915}.
The electron yield due to the plasmon excitations grows with decreasing the
projectile's energy (see also Figure~\ref{fig_Au_ElProd}).
On the contrary, the yield due to the atomic excitations exhibits a different
behavior.
The number of electrons emitted from the Au$_{32}$ cluster via the decay of
the atomic $5d$ excitation by a 0.3~MeV proton is larger than in the case of
a 1~MeV projectile.
However, a decrease of incident energy down to 0.1~MeV leads to an abrupt
decrease of the number of produced electrons.
As follows from Eq.~(\ref{ElScatter_to_PI}), at $\E_1 = 0.1$~MeV ($v_1 = 2.00073$~a.u.),
the dipole term of the $5d$ inelastic scattering cross section is strongly
suppressed, as the $\ln(v_1/\om R_{5d})$ term approaches zero.
In this case, the yield of electrons with kinetic energy below 5~eV due
to the plasmon excitation exceeds that due to the $5d$ atomic excitation by
the factor of about $10^3$.

%%%%%%%  Figure 12  %%%%%%%
\begin{figure*}[ht]
\centering
\includegraphics[width=0.9\textwidth,clip]{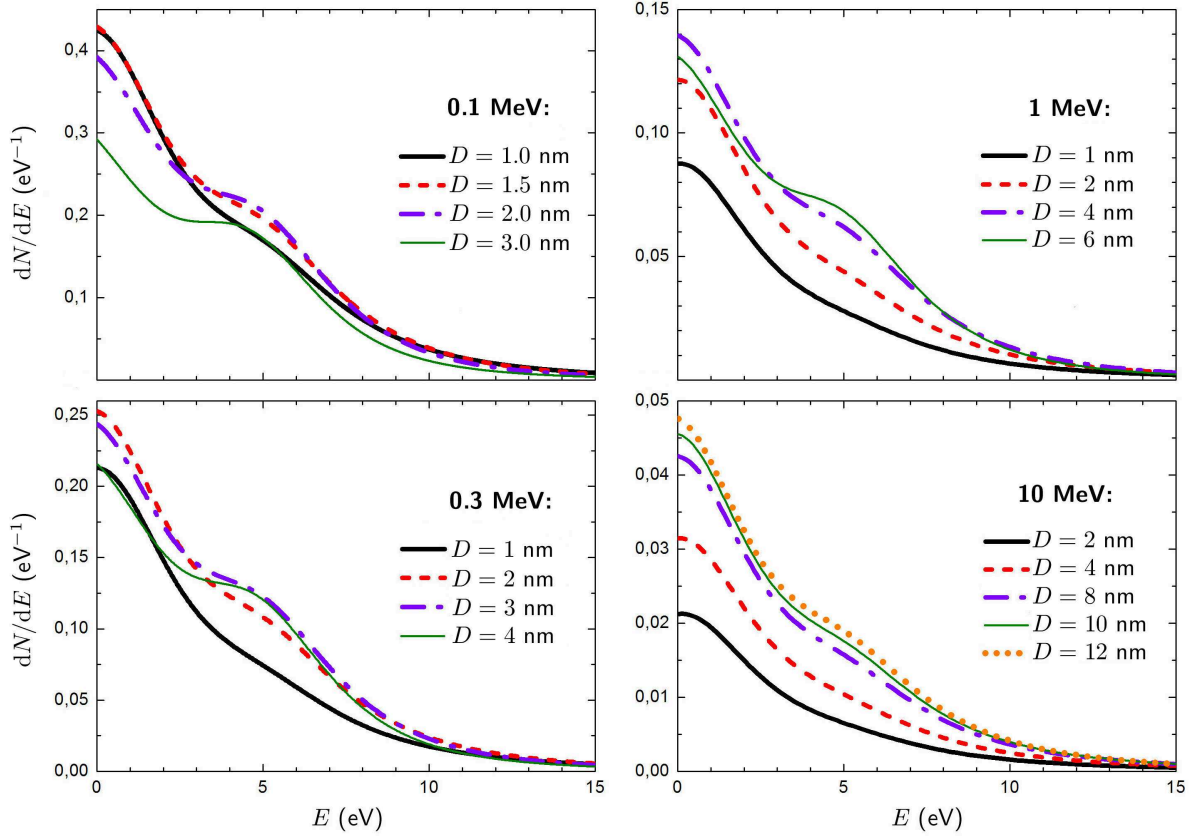}
\caption{
Number of electrons per unit energy produced via the plasmon excitation mechanism
in the solid gold nanoparticles of different size irradiated by the protons of
different kinetic energy.}
\label{fig_Au_ElProd_4}
\end{figure*}
%%%%%%%%%%%%%%%%%%%%%%%%%%%

Finally, let us analyze the role of the nanoparticle size on the intensity of
electron production.
In Figure~\ref{fig_Au_ElProd_4}, we consider the electron yield from the solid gold
nanoparticles of different size irradiated by the 0.1, 0.3, 1 and 10~MeV protons.
We focus on the systems of about $1 - 10$ nm in diameter.
Metal nanoparticles of this size range were studied recently in relation to the
radiotherapies with charged ions
\cite{Porcel_2010_Nanotechnology.21.085103, Porcel_2014_NanomedNanotechBiolMed}.
For the sake of clarity, we have calculated the number of electrons per unit energy,
${\rm d}N/{\rm d}E$.
At certain kinematic conditions, the contribution of the plasmon excitations saturates,
so that larger nanoparticles emit a smaller number of electrons via the plasmon damping
mechanism.

It was shown previously \cite{Gerchikov_1997_JPhysB.30.4133} that the dipole
mode of the plasmon excitations arising in a nanoparticle gives the dominating
contribution to the ionization cross section when the characteristic collision
%distance exceeds significantly the nanoparticle size, $v_1/\om \gg R$.
%, where $v_1$ is the incident projectile velocity.
distance exceeds significantly the nanoparticle size, $v_1/\om \gg D/2$
where $D$ is the nanoparticle diameter.
At large collision distances, the dipole contribution dominates over the
higher multipole contributions.
%, since the dipole potential decreases slower at large
%distances than the higher multipole potentials.
Terms with higher $l$ become significant only in the case when the collision
distances become comparable with the cluster size.
This means that for a given incident energy the plasmon mechanism of electron
production will be efficient for relatively small nanoparticles,
while the dipole plasmon mode will be suppressed for larger $D$.
This behavior is illustrated in Figure~\ref{fig_Au_ElProd_4}.
For instance, the number of low-energy electrons emitted via the surface plasmon
mechanism from a 2~nm nanoparticle irradiated by a 0.3~MeV proton
(dashed red curve) is higher than that from the 4~nm nanoparticle
(thin solid green curve).
In the former case, the characteristic collision distance
$v_1 / \om \approx 18$~a.u. becomes comparable with the nanoparticle radius,
$D/2 = 1~{\rm nm} \approx 19$~a.u.
Therefore, a larger nanoparticle with the diameter of 4~nm emits a smaller
number of electrons via the plasmon mechanism.
A small increase in the number of 5~eV electrons produced by the 4~nm nanoparticle
as compared to the smaller one is the result of an increased role of the volume
plasmon due to the increased volume/surface ratio.
A similar scenario holds for other incident velocities.
For a 1~MeV proton, the plasmon contribution to the electron yield saturates
%for the nanoparticle with the radius of approximately 2~nm, while for higher
for the nanoparticle with the diameter of approximately 4~nm, while for higher
energies ($\E_1$ = 10~MeV) the saturation takes place for the
%nanoparticle with radius of about 6~nm.
12~nm nanoparticle.

%higher multipole contributions become significant only in the case when the characteristic
%collision distances, $v / \om_r$, become comparable with the cluster size.
%In the opposite case of large collision distances, $v / \om_r \gg R$, the dipole contribution
%dominates over the higher multipole contributions, since the dipole potential decreases
%more slowly at large distances than the higher multipole potentials.
%The shape of the differential energy loss spectrum depends on the collision velocity, because
%the number of multipoles contributing significantly to the cross section is different in various
%collision energy ranges.
%This figure shows that the dipole resonance becomes dominant as the collision velocity increases.

%%%%%%%%%%%%%%%%%%%%%%%%%%%%%%%%%%%%%%%%%%%%%%%%%%%%%%%
\section{Conclusion}

In this work, we have performed a detailed theoretical and numerical analysis of
the electron production from gold nanoparticles due to irradiation by fast protons.
It has been demonstrated that due to the prominent collective response to an
external electric field, gold nanoparticles may significantly enhance the
yield of low-energy secondary electrons in the medium.
It has been shown that the significant increase in the number of emitted electrons
comes from the two distinct types of collective electron excitations, namely
plasmons and the excitation of $5d$ electrons in individual atoms of a nanoparticle.

The analysis of the plasmon contribution has been performed within the model
approach based on the plasmon resonance approximation.
To justify parameters of the model, photoabsorption spectra of several gold
nanoparticles have been calculated and compared with the spectra obtained by
means of time-dependent density-functional theory.
Our analysis has revealed that the broad feature positioned in the photoabsorption
spectra above 20~eV is related to the $5d$ giant resonance formed due to the
excitation of electrons in the $5d$ atomic shell.
The low-energy peak below 10~eV is caused by the formation of the surface plasmon,
which arises due to collective excitation of delocalized outer-shell valence electrons
in a whole cluster.
The integration of the oscillator strength in this energy region has revealed
that about 1.5 electrons from each atom contribute to this collective excitation.

The calculated yield of the electron production from gold nanoparticles has been
compared to that from pure water medium, based on the dielectric formalism.
It has been shown that the number of the low-energy electrons (with kinetic energy
of about a few electronvolts) produced by the gold nanoparticle of a given size
exceeds that produced by an equivalent volume of water by an order of magnitude.
At the energies of about several eV and higher, there is a prominent contribution
of the atomic $5d$ electrons to the ionization cross section and, subsequently,
to the electron production yield.
Accounting for the collective electron excitations leads to a significant increase
of the electron yield enhancement from the gold nanoparticle as compared to pure
water.
%Thus, the decay of the plasmon excitations formed in metallic nanoparticles is an
%important mechanism of generation of low-energy secondary electrons which should
%be considered along with other established mechanisms.
%By performing this analysis, we have stressed the sensitization effect of noble metal
%nanoparticles and their application in novel techniques of cancer therapy with ionizing
%radiation.
The importance of low-energy electrons is related to their ability to produce biodamage
by dissociative electron attachment.
The enhanced production of low-energy electrons will also lead to an increase
in the number of free radicals as well as other reactive species, like hydrogen
peroxide ${\rm H}_2{\rm O}_2$, which can travel the distances larger than the
cell nucleus \cite{Porcel_2014_NanomedNanotechBiolMed}.
Thus, these species can deliver damaging impacts onto the DNA from the radiation
induced damages associated with the presence of NPs in other cell compartments.

%%%%%%%%%%%%%%%%%%%%%%%%%%

The utilized approach for evaluation of the electron production yield has an
advantage as compared to other widely used approaches like track structure
Monte Carlo simulations, which are utilized in the microdosimetric calculations
(see Ref.~\cite{Waelzlein_2014_PhD_thesis} and references therein).
%
%The drawback of the Monte Carlo-based simulations is that they either do not allow
%one does not allow one to study the production of low-energy secondary electrons
%in any medium, except for the pure water.
%to account for the contribution of collective electron excitations or account for it
%incorrectly, utilizing impact ionization cross sections of isolated atoms as input
%parameters.
In most of the Monte Carlo simulations, the contribution of collective excitations,
which play a significant role in the ionization of gold and other noble metal
nanoparticles, is not accounted for.
An attempt to do it was made recently in Ref. \cite{Waelzlein_2014_PhysMedBiol.59.1441},
where the authors included the contribution of the volume plasmon excitation
when calculating the cross sections of electron and proton impact on noble metal
nanoparticles by means of Monte Carlo simulations.
It was stated that the plasmon excitation does not play an important role in the process
of electron emission from metallic nanoparticles, contributing much less to the overall
cross sections than individual excitations.
In this paper, we have demonstrated that the relative contribution of the surface
collective electron excitation to the cross section exceeds that of the volume plasmon
by an order of magnitude.
Thus, the leading mechanism of electron production by gold nanoparticles is related
to the surface plasmon.
As a result of this analysis, we have demonstrated that the emission of low-energy
electrons from the gold nanoparticles is a prominent effect, which should be accounted
for when estimating the production of secondary electrons in a biological medium with
embedded nanoparticles.

%%%%%%%%%%%%%%%%%%%%%%%%%%

The calculated spectra of emitted electrons can further be used as the input data
for the multiscale approach to the physics of radiation damage
\cite{Surdutovich_2014_EPJD_Colloquia_Paper}.
This approach has the goal of developing knowledge about biodamage at the nanoscale
and molecular level and finding the relation between the characteristics of incident
particles and the resultant damage
\cite{Surdutovich_2014_EPJD_Colloquia_Paper, Solov'yov_2009_PhysRevE.79.011909}.
Despite the fact that we have explored only the case of gold nanoparticles,
we have introduced a general methodology, which can also be applied for other
nanoscale systems, currently proposed as sensitizers in cancer therapy.
The methodology can also be extended for collisions with heavier ions.
In this case, the cross section obtained should be multiplied by a $Z^2$
scaling factor due to the change of Coulomb field of the projectile.
Thus, the presented methodology can be utilized for studying irradiation of
sensitizing nanoparticles by carbon ions, which are the most clinically used
projectiles, besides protons.

%%%%%%%%%%%%%%%%%%%%%%%%%%%%%%%%%%%%%%%%%%%%%%%%%%%%%%%
\section*{Acknowledgements}

We are grateful to Pablo de Vera for providing us the data on electron production
in pure water medium.
We also acknowledge the Frankfurt Center for Scientific Computing for the opportunity
of carrying out resource-demanding DFT calculations.
%
%\textcolor{red}{
%We also thank the anonymous referees for constructive criticism.
%}

%%%%%%%%%%%%%%%%%%%%%%%%%%%%%%%%%%%%%%%%%%%%%%%%%%%%%%%
%\section*{References}

\end{document}